\documentclass[aps,prb,unsortedaddress,superscriptaddress,showpacs,twocolumn,floatfix]{revtex4-1}

\usepackage{amsmath}
\usepackage{graphicx}

\begin{document}

\title{Longitudinal muon spin relaxation in high purity aluminum and silver}

\affiliation{University of British Columbia, Vancouver, British Columbia, V6T 1Z1, Canada}
\affiliation{TRIUMF, Vancouver, British Columbia, V6T 2A3, Canada}
\affiliation{University of Regina, Regina, Saskatchewan, S4S 0A2, Canada}
\affiliation{Kurchatov Institute, Moscow, 123182, Russia}

\author{J.F.~Bueno}
\affiliation{University of British Columbia, Vancouver, British Columbia, V6T 1Z1, Canada}

\author{D.J.~Arseneau}
\affiliation{TRIUMF, Vancouver, British Columbia, V6T 2A3, Canada}

\author{R.~Bayes}
\altaffiliation[Present Address: ]{School of Physics and Astronomy, University of Glasgow, Glasgow, G12 8QQ, Scotland}
\affiliation{TRIUMF, Vancouver, British Columbia, V6T 2A3, Canada}

\author{J.H.~Brewer}
\affiliation{University of British Columbia, Vancouver, British Columbia, V6T 1Z1, Canada}

\author{W.~Faszer}
\affiliation{TRIUMF, Vancouver, British Columbia, V6T 2A3, Canada}

\author{M.D.~Hasinoff}
\affiliation{University of British Columbia, Vancouver, British Columbia, V6T 1Z1, Canada}

\author{G.M.~Marshall}
\affiliation{TRIUMF, Vancouver, British Columbia, V6T 2A3, Canada}

\author{E.L.~Mathie}
\affiliation{University of Regina, Regina, Saskatchewan, S4S 0A2, Canada}

\author{R.E.~Mischke}
\email{mischke@triumf.ca}
\affiliation{TRIUMF, Vancouver, British Columbia, V6T 2A3, Canada}

\author{G.D.~Morris}
\affiliation{TRIUMF, Vancouver, British Columbia, V6T 2A3, Canada}

\author{K.~Olchanski}
\affiliation{TRIUMF, Vancouver, British Columbia, V6T 2A3, Canada}

\author{V.~Selivanov}
\affiliation{Kurchatov Institute, Moscow, 123182, Russia}

\author{R.~Tacik}
\affiliation{University of Regina, Regina, Saskatchewan, S4S 0A2, Canada}

\date{\today}

\newcommand{\pmupi} {\ensuremath{P_\mu^\pi}\,}
\newcommand{\pmuxi} {\ensuremath{\pmupi\xi}}
\newcommand{\cm}    {\,\textrm{cm}}
\newcommand{\mrad}  {\,\textrm{mrad}}
\newcommand{\tes}   {\,\textrm{T}}
\newcommand{\MeV}   {\,\textrm{MeV}}
\newcommand{\MeVc}  {\, \textrm{MeV}/\textrm{c}}
\newcommand{\GeVcc} {\,\textrm{GeV}/\textrm{c}^2}
\newcommand{\mum}   {\, \mu \textrm{m}}
\newcommand{\mus}   {\, \mu \textrm{s}}
\newcommand{\ns}    {\,\textrm{ns}}
\newcommand{\mr}    {\, \textrm{mrad}}
\newcommand{\abs}[1]{\ensuremath{\left\lvert#1\right\rvert}} 
\newcommand{\ct}    {\ensuremath{\cos\theta}}
\newcommand{\musr}  {$\mu^+$SR}
\newcommand{\pms}   {\ensuremath{\,\textrm{ms}^{-1}}}
\renewcommand{\deg} {\ensuremath{^\circ}}  

\begin{abstract}
The time dependence of muon spin relaxation has been measured in high purity 
aluminum and silver samples in a longitudinal 2 T magnetic field at room 
temperature, using time-differential \musr. For times greater than 10 ns,
the shape fits well to a single exponential with relaxation rates of
$\lambda_{\textrm{Al}} = 1.3 \pm 0.2\,(\textrm{stat.}) \pm 0.3\,(\textrm{syst.})\,\pms$
and $\lambda_{\textrm{Ag}} = 1.0 \pm 0.2\,(\textrm{stat.}) \pm 0.2\,(\textrm{syst.})\,\pms$.
\end{abstract}

\pacs{76.75.+i}

\maketitle

\section{Introduction}
\label{s:intro}
When positive muons are stopped in a high purity metal with the muon 
spin along the direction of an applied magnetic field,
the muons can be depolarized by interactions with nuclear dipole moments,
conduction electrons, and paramagnetic impurities.
The form of the resulting depolarization has been studied using the \musr\
technique,\cite{Brewer:1994} but has not been 
quantified to the level needed for the  
TWIST experiment,\cite{TWIST} in which the absolute polarization of
the muons must be known with high precision. 
TWIST used stopping targets of aluminum and 
silver with purity greater than $99.999\%$ immersed in an external
2.0 T longitudinal magnetic field at room temperature.
Because the TWIST drift chamber detector system recorded
ionization from incoming muons as well as positrons with a
relatively slow drift gas, the analysis discarded decay positrons
within $1.05\mus$ of the muon stopping time due to possible time
overlap of muon and positron ionization.
Therefore, a time differential \musr\ experiment was undertaken to 
make a precise measurement of the relaxation rate, especially
for decay times below $1.05\mus$.

\section{Depolarization mechanisms for stopped muons}
\label{s:stationary_time_dependent_depolarization} 

After motional thermalization, positive muons are limited
to interstitial and substitutional sites.
When nearly thermalized, a muon in a good metal attracts   
a screening charge of conduction electrons.\cite{Cox:1987}
At room temperature, a muon is usually diffusing between energetically
allowed sites before decaying.  
The conduction electrons in aluminum
and silver efficiently screen the ionic potentials 
over all distances greater than the Fermi-Thomas screening length.
 
Muons can become trapped at crystal defects, 
of which there are a wide variety; the most common originate
from the manufacturing process, such as when a metal
is cold-rolled to produce a thin foil.  
Trapping by such defects 
 makes the muon mobility strongly sample dependent.
The defects are enhanced by quenching, and can be diminished by annealing.

In slowing down from high energy, the muon itself can 
cause lattice defects
as it imparts recoil energy to 
lattice ions in $\sim 10^{-17}\,\textrm{s}$ 
and the lattice distributes this energy to neighbouring atoms 
in $\sim 10^{-12}\,\textrm{s}$.\cite{Schilling:1978}
A nucleus can be knocked out of its lattice position
into an interstitial site, leaving a vacancy (Frenkel defect). 
However, these vacancies are unlikely to affect 
the muon's diffusion, since 
the muon thermalizes $\sim 1 \; \mum$ from
the last defect introduced.\cite{Brice:1978}  

The magnetic field experienced by a stationary muon 
due to the magnetic dipole moments of 
nuclei and lattice impurities can be modelled 
as static, isotropic, and Gaussian.  
If the muon samples a new field taken at random from the same distribution 
every time it ``hops'' to a new lattice site, 
the resulting depolarization is given by\cite{Dalmas:1997} 
\begin{equation}
  P_{\mu}(t) = P_{\mu}(0) \exp{\left\{ -\frac{2 \Delta^2}{\nu^2} 
   \left[ \exp(-\nu t) -1 + \nu t \right] \right\} },
\label{e:dalmas_equation}
\end{equation}
and is valid for $\nu/\Delta$ sufficiently large, where $\Delta$ is a measure of
the magnetic field distribution 
and $1/\nu$ is the mean time between muon hops.  
Each field component is presumed to have an independent Gaussian distribution 
$ \mathcal{D}(B_{\textrm{local}}) \sim \exp\left[ 
 - B_{\textrm{local}}^2/(2\Delta^2/\gamma_{\mu}^2) \right]$, 
where $\gamma_\mu$ is the muon's gyromagnetic ratio 
and $\Delta/\gamma_\mu$ is the standard deviation of a Gaussian distribution 
of magnetic fields.  

If an external field $\vec{B}_{\textrm{ext}}$ is now applied in a direction
\emph{transverse\/} to the muon polarization, the muon spins precess and are 
depolarized according to the Abragam formula:\cite{Abragam:1986}
\begin{equation}
  P_{\mu}(t) = P_{\mu}(0) \exp{\left\{ 
   -\frac{\Delta^2}{\nu^2} \left[ \exp(-\nu t) -1 + \nu t \right] 
  \right\} } \cos{(\omega_\mu t)} \; ,
\label{e:abragam_formula}
\end{equation}
where $\omega_\mu = \gamma_\mu B_{\textrm{ext}}$.  
In the ``motional narrowing'' limit, the muons move 
quickly so that $\nu$ is large, $\exp{(-\nu t)} \to 0$
and the {\sl envelope\/} of Eq.~\eqref{e:abragam_formula} approaches an
exponential time dependence.  In the static limit, 
the envelope instead approaches a Gaussian time dependence.  

If a \emph{longitudinal\/} field $B_0$ is applied, 
the static relaxation function becomes \cite{Hayano:1979}
\begin{eqnarray}
  P_{\mu}(t) & = & 1 - \frac{2 \Delta^2}{\omega_0^2} 
  \left[ 1 - \exp{\left(-\tfrac{1}{2}\Delta^2 t^2\right)} \cos{\omega_0 t}\right] 
 \nonumber\\
  &   & + \frac{2 \Delta^4}{\omega_0^3} 
 \int_0^t{\exp{\left(-\frac{1}{2}\Delta^2 \tau^2\right)} 
  \sin{\omega_0 \tau} \, d\tau} \; , 
\label{e:longitudinal_blah}
\end{eqnarray}
where $\omega_0 = \gamma_\mu B_0 \approx 272$ MHz, and the longitudinal
field is seen to suppress the depolarization due to nuclear
dipole moments.  
The largest observed nuclear dipolar field on a muon in a crystal
is $\Delta/\gamma_\mu = 0.47$ mT,\cite{Dalmas:1992}
while the applied longitudinal field in this experiment was $B_0 = 2$~T, 
so that $(2 \Delta^2/\omega_0^2) \lesssim 10^{-7}$. Of course, in
our experiment the muons are not static.  The cumlative loss of
polarization increases with increasing $\nu$ as long as 
$\nu \ll \omega_0$. For $\nu \gg \omega_0$ 
``motional averaging'' sets in and the relaxation
rate goes down again; this is known as the ``$T_1$ minimum''
effect.\cite{Slichter:1978}  So the fastest relaxation would be 
for $\nu \sim 100 \, \mu\textrm{s}^{-1}$; assuming muons lose up to $10^{-7}$
of their polarization at each hop, in that case the relaxation rate is 
$<0.01 \, \textrm{ms}^{-1}$. At room temperature in 
aluminum\cite{Richter:1986} 
$\nu \approx 10^{11} \, \textrm{s}^{-1}$, which is far past the ``$T_1$ minimum''.
Depolarization by nuclear dipole moments 
is therefore negligible for this experiment.

Korringa relaxation\cite{Korringa:1950} caused by 
a hyperfine contact interaction between the muon spin and 
fluctuating conduction electron spins 
is the most likely cause of any muon depolarization in this experiment.  
The net hyperfine coupling is an average over all 
electron spin orientations.\cite{Blundell:2001,Brewer:1975}  
The resulting exponential relaxation rate is given by
\begin{equation}
P_\mu(t) = P_\mu(0)\exp(-\lambda t).
\label{e:pmut_relax}
\end{equation}
where $\lambda$ is proportional to $K_\mu^2 T$. The 
muon Knight shift $K_\mu = (\omega - \omega_0)/\omega_0$  
is the fractional muon frequency shift due to the same hyperfine interaction.  
Korringa relaxation has been observed in
several non-magnetic metals (lead, cadmium, zinc, copper),
where the muon relaxation rates increase linearly with temperature 
and are robust to field changes in the range 0.01~T to 0.20~T,\cite{Cox:2000} 
as expected.\cite{Blundell:2001}  

As we have seen, relaxation of muons by nuclear dipole moments 
is heavily suppressed by a longitudinal magnetic field.  
The electronic dipole moments of paramagnetic ions 
are much larger and can produce fields up to 0.1~T 
at a distance of one lattice spacing.\cite{Brown:1981,Blairpaper} 
Thus they are much more difficult to decouple from the muon 
and may cause exponential relaxation of muons 
in a crystal with significant concentrations of 
paramagnetic impurities.\cite{Brown:1981,Abragam:1986,Heffner:1981}

In summary, muon spin relaxation due to nuclear dipole moments 
exhibits an exponential form 
when the muons hop rapidly, as expected at room temperature.  
Non-exponential static dipolar relaxation may occur 
if the muons become trapped at defects or vacancies, 
which is possible in our aluminum foil since it was not annealed, 
but any such static relaxation due to nuclear dipole moments 
is heavily suppressed by the presence 
of a strong longitudinal magnetic field.
In the silver foil, trapping at paramagnetic impurities cannot be excluded.  
The appropriate form is still
exponential, as long as the muons
diffuse sufficiently fast to find the impurities promptly.  
This may be the case in an annealed sample 
such as our silver foil. For both foils, Korringa relaxation is expected 
to be the dominant mechanism for depolarization.  

\section{Previous measurements}
We were unable to find any measurements from \musr\ taken under conditions
comparable to those of TWIST. 
Depolarization in aluminum and silver from nuclear
dipole moments has been measured in \musr\ experiments,
using a transverse magnetic field arrangement.
There are more studies on aluminum since its nuclear dipole
moment is about $35$ times larger than silver.
Even with its large dipole moment, high purity
aluminum leads to almost negligible depolarization
down to 1 K.\cite{Kossler:1978,Hartmann:1978,Kehr:1982,Kehr:1984,Hartmann:1988}
Consequently, a measurable depolarization is typically seen only in samples 
that have been doped with impurities.
\cite{Hartmann:1990,Cox:1987,Hartmann:1980}

There is a contradictory measurement,\cite{Stoker:1985_2}
which used aluminum and silver targets of $99.99\%$ purity
in a transverse field arrangement, at room temperature, and observed
a Gaussian form for the depolarization in aluminum. They explained
this anomalous result as due to muons trapping
in defects, which originated from the cold-rolling
during manufacture of the foil.\cite{private_stoker}

The result from an intermediate phase of TWIST,\cite{Blairpaper} which used 
a 2 T longitudinal magnetic field, is 
$\lambda_{\textrm{Al}} = 1.6 \pm 0.3\,\pms$. One other 
measurement\cite{Jodidio} was available prior to this experiment,
where a high purity aluminum sample was measured in a 1.1 T
longitudinal field. The result, based on four points in the time range of
0.1 to 10 $\mu$s, is 
$\lambda_{\textrm{Al}} = 0.43 \pm 0.34\,\pms$.
It is difficult to compare the results from these two experiments due
to differences in magnetic fields, limited statistical precision, 
and suspected systematic uncertainties in determining $\lambda$. 
Comparing results between different experiments has the additional
complication that if the depolarization is dominated by small
impurities and/or defects introduced during manufacturing, 
then the result is expected to be dependent on the
sheet from which the sample is taken.

\section{Experimental Details}

\begin{figure}[!hbt]
  \begin{center}
    \includegraphics[width=3.4in]{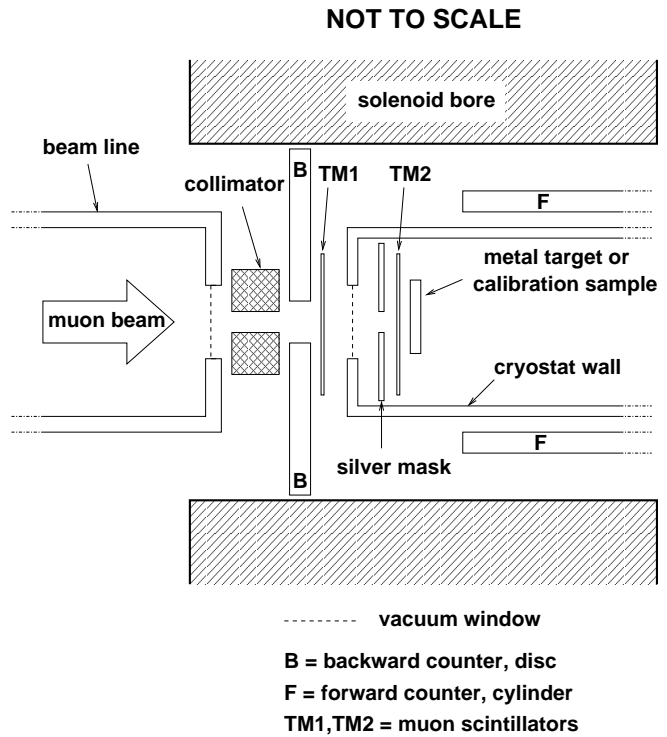}
  \end{center}
  \caption{Schematic for the experiment.
    A collimated muon beam enters from the left and muon decays are recorded
    in one set of histograms if they are detected only by the first 
    scintillator (TM1), stopping mostly in the Ag mask, or in another set of
    histograms if they are recorded by both
    scintillators (TM1 and TM2), mostly stopping in the target.
    The backward positron counter (B) is protected from
    the muon beam by a collimator. Materials such as the
    photomultiplier tubes and light guides are not shown.}
  \label{f:musr_schematic}
\end{figure}

In December 2006 we acquired \musr\ data  using the 
facilities of the TRIUMF Centre for Molecular \& Materials Science.
The M20B beamline, which was
a dedicated surface muon channel,\cite{Beveridge:1985} included
collimators for the muons and a DC separator that
removed positrons by velocity selection. 
The separator also rotated the muon spin by $11\deg$, creating a small 
transverse polarization component and reducing the
longitudinal polarization by 2\%. The muon rate was typically 20-40 kHz.
The muon target and positron counters are shown schematically
in Fig.~\ref{f:musr_schematic}.
Muons could stop in the sample under investigation,
one of the scintillators, or an Ag mask of
purity $99.99\%$. Most data were taken with a TM2 scintillator
of nominal thickness $254 \mum$, but data were also taken with
nominal thicknesses of $127$ and $508 \mum$.
The backward (B) and forward (F) positron counters were
approximately $0.6\cm$ thick. The backward counter was
a disc of about $8.0\cm$ diameter, with a $2.5\cm$
hole for the muon beam. The F counter was tubular, with inner 
diameter of $10.5\cm$ and length $35.5\cm$.
The target module was placed in the
HELIOS superconducting solenoid that supplied
a longitudinal $2\tes$ magnetic field over
the stopping target.
The target was at room temperature, and the region around the target was
evacuated.

The aluminum stopping target was purchased from Goodfellow Corporation,\cite{Goodfellow} 
who specified the typical impurities as 0.3 ppm of Cu,
0.3 ppm of Fe, 1.2 ppm of Mg, and 0.8 ppm of Si. 
In aluminum, the muons are not trapped by impurities
above $\approx$100 K.\cite{Hartmann:1988}
High purity aluminum has been
studied under annealing and quenching,\cite{Gauster:1977} over a temperature
range of 19 to 900 K:
most defects were found to be absent after allowing the
quenched sample to reach room temperature.
The silver stopping target
was purchased from ESPI Metals,\cite{ESPI} who specified the typical impurities as
2 ppm of Fe, $<$ 2 ppm of Bi, 0.6 ppm of Cu, and 0.6 ppm of Pd.
It was annealed in an inert argon atmosphere, after machining.
In silver, there is evidence that room temperature trapping
at impurities can occur,\cite{Heffner:1981} but that would not be of 
consequence for our annealed sample.

The data acquisition system recorded the positron time of
arrival relative to that of the muon, in two pairs of 
histograms, each with $19200$ channels of width $0.78125\ns$.
Data were recorded for the B and F counters, in parallel for muons stopping in the Ag mask and
for those stopping in the target, separated by the absence or presence
of a signal in TM2 (see Fig.~\ref{f:musr_schematic}).
There was a background of muon decays in or near the collimator,
scattered positrons, and muon decays
between the muon detector and the target.
The electronics removed the background from muon
pile-up, where a second muon enters the sample
region during the same $15 \mus$ long data gate as the original muon.
The positron gate was open for $1\mus$
before the muon trigger, so that events were recorded 
that could not possibly be caused by decay of the detected muon.
The aim was to fix the background to the average
of the number of decays recorded in this  $1\mus$ interval.

The stopping targets are listed in Table \ref{t:musr_data}.
For sets E-K, the target consisted of two layers, one of metal from
the same foils as the TWIST targets, the other a calibration 
sample (CS). The orientation of the layers could be reversed to expose
either one to the stopping muon beam.
The metal portion
was either 10 folded layers of Al or
6 folded layers of Ag, both of which were sufficiently
thick to stop all the muons in the surface muon beam. The layers were bound
together with $3.6\mum$ thickness Mylar.
The CS was a 2.5 cm diameter, 0.2 cm thick (mass $4.6 \textrm{g}$) pressed powder disc of 
Gd$_2$Ti$_2$O$_7$, a geometrically frustrated antiferromagnetic pyrochlore.
At room temperature Gd$_2$Ti$_2$O$_7$ is a paramagnet in which fluctuating 
magnetic moments are known to depolarize muon spins within a few microseconds,
with an exponential form of $P_{\mu}(t)$.\cite{Dunsiger}
Note that no grease or glue was used in
the assembly of the targets.
Sets A-D used targets of pure CS or pure metal,
before we realized the importance of maintaining
constant material thickness in the decay positron path when making
systematic comparisons of decay asymmetry.
We anticipated that determining the fraction of muons stopping in the
trigger scintillator (TM2) would be a dominant source of systematic
uncertainty, thus for sets I-K the thickness of the trigger scintillator 
was altered.

\begin{table}[!hbt]
  \caption{Data sets for the \musr\ experiment.
  Sets C and D had a low muon rate.
  Set D had the DC separator on a high setting. In the target
  column, the upstream material of the reversible target is listed first.}
  \label{t:musr_data}
  \begin{tabular*}{0.95\columnwidth}{@{\extracolsep{\fill}}cccccc}
  \hline\hline
  Data set  &  Target  &  Scintillator & Duration   &  \multicolumn{2}{c}{$\mu^+$ counts} \\
            &          &  thickness    & (hours)    &  Ag mask          &  sample         \\
            &          &  ($\mum$)     &            &  ($\times 10^9$)  & ($\times 10^9$) \\
  \hline
  A         &  CS\footnotemark[1]& 254 & 14.1       & 1.2               & 0.8  \\
  B         &  Al            &     254 & 8.0        & 0.7               & 0.5  \\
  C         &  Al            &     254 & 15.7       & 0.7               & 0.5  \\
  D         &  Al            &     254 & 5.7        & 0.3               & 0.2  \\
  E         &  CS+Al         &  254    & 24.0       & 1.7               & 1.2  \\
  F         &  Al+CS         &  254    & 40.1       & 2.9               & 2.0  \\
  G         &  Ag+CS         & 254     & 32.4      & 2.5               & 1.7 \\
  H         &  CS+Ag         & 254     & 22.8      & 1.8               & 1.2 \\
  I         &  CS+Ag         & 508     & 18.0      & 1.3               & 0.9 \\
  J         &  CS+Ag         & 127     & 22.2      & 1.5               & 1.0 \\
  K         &  Ag+CS         & 127     & 13.3      & 1.0               & 0.7 \\
  \hline\hline
  \end{tabular*}
  \footnotetext[1]{CS: Gd$_2$Ti$_2$O$_7$ calibration sample}
\end{table}

\section{Analysis}

When analyzed with a Fourier transform algorithm, the data showed time 
structures with periods of $43.37\ns$ and $3.68\ns$,
corresponding to the period of the TRIUMF cyclotron
and the precession frequency of the muon transverse
polarization components in the
$2.0\tes$ magnetic field.
To minimize the impact of possible effects, we rebinned our
data using time bins of width $89.84\ns$ starting $10\ns$ after
the muon stop time,
and we later verified that our results are robust to the
binning used.

\begin{figure}[!hbt]
  \includegraphics[width=\columnwidth]{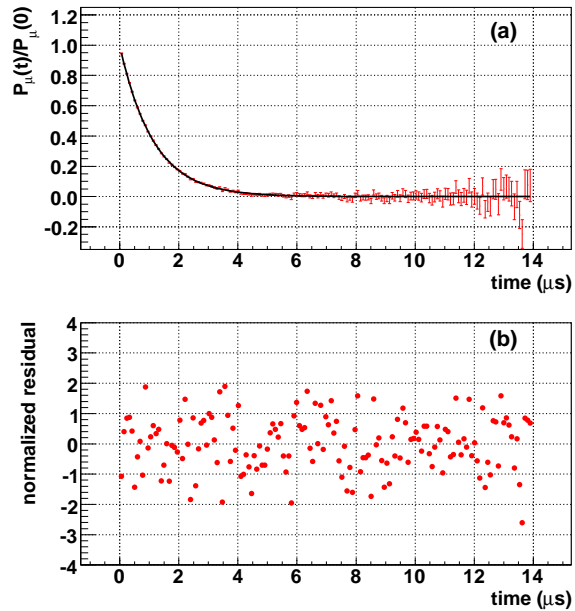}
  \caption{Depolarization for the CS (Gd$_2$Ti$_2$O$_7$).
   (a) relative polarization and (b) normalized residuals. The fit is 
   $P_\mu(t)/P_\mu(0) = \exp(-\lambda t)$ to data from set E, with
   $\lambda = (0.860 \pm 0.004) \mus^{-1}$ and the $\chi^2/\textrm{ndf} = 285.5/302$.}
  \label{spin_glass}
\end{figure}

We simultaneously fit the B and F decay count histograms
with standard coupled \musr\ equations,
\begin{eqnarray}
  n_b \left(t\right) & = &
  b_b +   N_0 e^{-t/\tau_\mu} \left[ 1 + A_b P_\mu \left(t\right) \right]
  \label{eq:finalmusrnb} \\
  n_f \left(t\right) & = &
  b_f + r N_0 e^{-t/\tau_\mu} \left[ 1 + A_f P_\mu \left(t\right) \right],
  \label{eq:finalmusrnf}
\end{eqnarray}
where $n_b$ and $n_f$ are the number of B and F
counts, $b_b$ and $b_f$ are the backgrounds in each counter,
$\tau_\mu$ = 2.19703 $\mu$s is the muon lifetime, $N_0$ is the histogram normalization,
$A_b$ and $A_f$ are the empirical asymmetries that depend only on
the counter geometry and the material through which the decay positrons
pass. The parameter $r$ is introduced to account for
differences between the B and F counters' energy and
angular acceptance. The CS sample was fit first, using
three forms for the depolarization:
a single exponential, $P_\mu(t) = P_\mu(0) \exp(-\lambda t)$, 
a sum of two exponentials, and a power law, $P_\mu(t) = \exp(-at^p)$.
The data were consistent with a single exponential relaxation,
so this form is assumed in the following. (See Fig.~\ref{spin_glass}.)

The backgrounds in each counter were determined using the $1\mus$ of 
pre-trigger data, and also by leaving the background as a free fit parameter. 
We found that the two measurements differed by more than three standard 
devations, probably due to contamination of the background from long-lived 
muons. Since the backgrounds could be well determined using the decay data 
and had a weak correlation with the other fit parameters, we chose to fit the
background level with a free parameter.

\section{Results and Systematic Uncertainties}

The parameters from fits to data for the Ag mask, which only required a trigger
in TM1, were affected by changes in
the target and running conditions.  This dependence was seen dominantly in 
the values of the parameter $r$. The value of $\lambda$ was somewhat sensitive
to the target, but the average measured
value of $\lambda_{\textrm{mask}} = (1.07 \pm 0.07) \textrm{ms}^{-1}$ agrees
with the result for the high purity Ag given below. Systematic errors
were not evaluated for the Ag mask.

We used the data from sets I and J to determine
the relaxation rate for muons stopping in the
trigger scintillator.
Initially the data with a $127\mum$ thickness scintillator 
were fit assuming that all muons passed through the scintillator and stopped in
the CS. With the CS relaxation fixed, we fit a sum of 
exponentials to the data with scintillator thickness $508 \mum$, which 
provided an estimate of the relaxation within the scintillator itself.
This process was iterated, and we found that
a single exponential is adequate to describe
the scintillator depolarization, with
$86.0 \pm 0.3 \%$ of the muons stopping in the $508\mum$ thickness 
scintillator and $\lambda_{\textrm{scint}} = (0.0132 \pm 0.0008) \mus^{-1}$.
Then the CS relaxation
is $\lambda_{\textrm{CS}} = (0.866 \pm 0.008) \mus^{-1}$. 
With the relaxation rate
for the scintillator component fixed,
an attempt was made to determine the fraction of muons
stopping in the other two scintillators. For the $127\mum$
thickness scintillator we find $(4 \pm 9)\%$, and
for the $254\mum$ scintillator we find $(0 \pm 10)\%$ for set E
and $(8 \pm 8)\%$ for set H. These results are
clearly imprecise, so we resorted to an evaluation of the
fraction using a SRIM\cite{SRIM} simulation of the muon stopping position
distribution.
Based on estimates of the M20B beam line mean momentum (between 29.2 and
28.7 MeV/c) and
momentum resolution (between 1 and 2\%), and of materials in the path of the stopping
muons, the results of the simulation showed that $(3 \pm 3)\%$
of the muons could have stopped in the $254\mum$
scintillator, and no more than $0.2\%$ 
in the $127\mum$ scintillator.
We correct our result
assuming that $3\%$ of muons stop and depolarize
in the $254\mum$ TM2 trigger scintillator rather than
the metal target, and assign a large systematic
uncertainty that allows for between $1\%$ and $6\%$
of muons stopping in the scintillator.

\begin{figure}[!hbt]
  \includegraphics[width=\columnwidth]{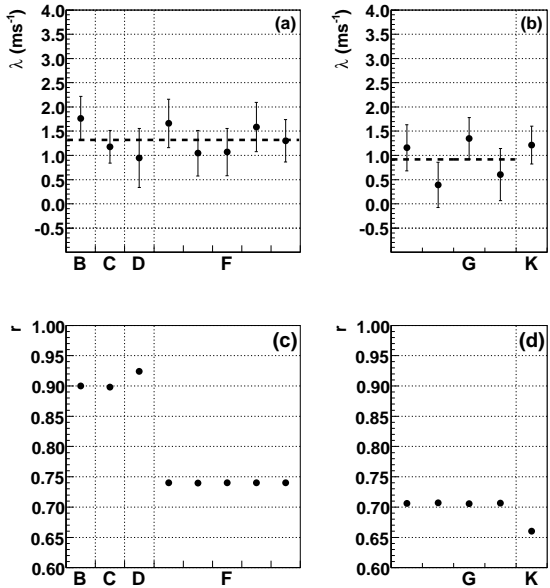}
  \caption{Results for fits to the metal samples. (a) $\lambda$ for 
  Al, (b) $\lambda$ for Ag, (c) $r$ for Al, and (d) $r$ for Ag. 
  Sets F and G had multiple runs.
  The parameter $r$ is highly sensitive to target
  placement, which occurred at the beginning of
  each set.}
  \label{f:musr_results}
\end{figure}

The empirical coefficients ($A_b$ and $A_f$)
from the CS analysis were determined with statistical
precision $<0.3\%$ and were constrained to be $A_b = 0.185$
and $A_f = -0.238$ while fitting the asymmetry data for the metals.
There is a negligible contribution to the systematic uncertainty from the fixed
$A_b$ and $A_f$ values; a change of $10\%$ in either of these parameters 
altered the relaxation rate by at most $0.12\pms$.
The $r$ parameter was left free since it was found to
be highly sensitive to the exact placement of the
target. This is shown in Fig.~\ref{f:musr_results},
which also includes the results for the
relaxation parameter $\lambda$ in Eq.~\eqref{e:pmut_relax}.

A significant time structure of the background
would give rise to a systematic uncertainty.  Because the confidence
levels for the fits were reasonable, except for one set that
still produced a consistent relaxation rate, there was no need to
investigate a possible time structure of the background. Also, the sets 
with different muon rates and DC
separator settings showed no evidence
of an inconsistent relaxation rate.
Therefore no systematic uncertainties
are assigned for these effects.

The uncertainty in the fraction of muons stopping in the $254\mum$ scintillator
is much larger than anticipated,
and dominates all others. Unfortunately we cannot
reduce this uncertainty, since a precise
measurement of the M20B mean momentum and resolution
was not taken with our beamline settings.

Our results for the $254\mum$ scintillator are then
\begin{eqnarray}
  \lambda_{\textrm{Al}} &=& 1.3 \pm 0.2\,(\textrm{stat.}) \pm 0.3\,(\textrm{syst.})\,\pms \textrm{and} \\
  \lambda_{\textrm{Ag}} &=& 0.9 \pm 0.2\,(\textrm{stat.}) \pm 0.2\,(\textrm{syst.})\,\pms.
\end{eqnarray}

\begin{figure}[hbt]
  \includegraphics[width=\columnwidth]{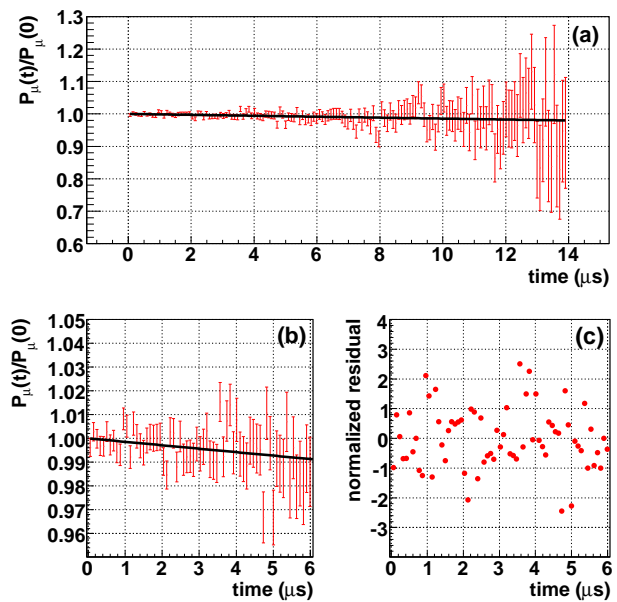}
  \caption{Aluminum relative polarization vs time. (a) full time range,
           (b) first 6 $\mu$s, and (c) normalized residuals from fit.
           The fit is $P_\mu(t)/P_\mu(0) = \exp(-\lambda t)$ to one run of set F, with
           $\lambda = (1.5 \pm 0.5) \pms$ and the $\chi^2/\textrm{ndf} = 280.6/304$.}
  \label{f:asymmetry_fit_2517}
\end{figure}

\begin{figure}[hbt]
  \includegraphics[width=\columnwidth]{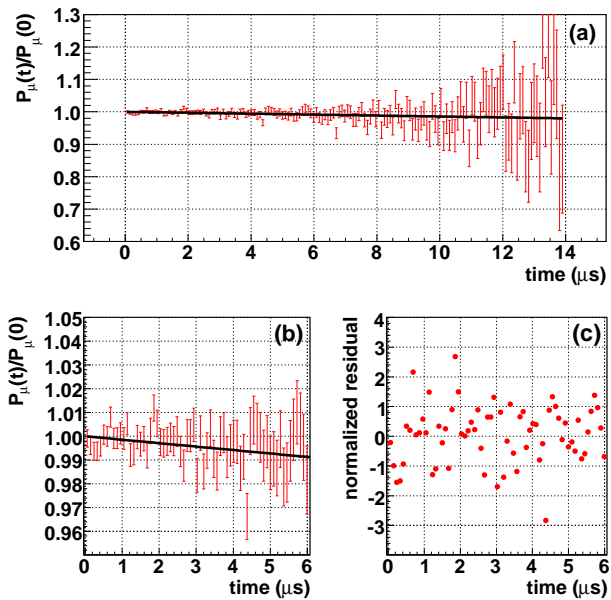}
  \caption{Silver relative polarization vs time. (a) full time range,
           (b) first 6 $\mu$s, and (c) normalized residuals from fit.
           The fit is $P_\mu(t)/P_\mu(0) = \exp(-\lambda t)$ to one run of set G, with
           $\lambda = (1.5 \pm 0.5) \pms$ and the $\chi^2/\textrm{ndf} = 300.2/304$.}
  \label{f:asymmetry_fit_2522}
\end{figure}

A single exponential fit is shown for an aluminum run
in Fig.~\ref{f:asymmetry_fit_2517} and for a silver run in Fig.~\ref{f:asymmetry_fit_2522}.
There is no evidence that anything beyond a single exponential is
needed to describe the depolarization.
There is no observable fast depolarization component below $1\mus$.

For the single run with the thin ($127\mum$) scintillator,
the contribution from scintillator stops is negligible,
but the statistical uncertainty dominates so that
\begin{equation}
  \lambda_{\textrm{Ag}} = 1.2 \pm 0.4\,(\textrm{stat.})\,\pms.
\end{equation}
Combining the two results for silver yields
\begin{equation}
  \lambda_{\textrm{Ag}} = 1.0 \pm 0.2\,(\textrm{stat.}) \pm 0.2\,(\textrm{syst.})\,\pms.
\end{equation}

\section{Conclusions}

The measured
relaxation rates for silver and
aluminum differ by less than  a factor of two, yet the
nuclear dipole moments differ by a factor
of 35, providing further evidence that
the depolarization is not from nuclear
dipole moments.
These results are consistent with those from the final phase of the
TWIST experiment,\cite{Bueno} but with somewhat larger uncertainties.
The unique conclusion from this experiment
is that no additional depolarization components
exist in the time range $0.010 < t < 1.000\mus$; this is very
important for the interpretation of TWIST data.
We are unaware of any experimental technique that
would allow us to readily study the depolarization below
$10\ns$, but we are also unaware of any credible
models for muon depolarization within the
first $10\ns$ in nonmagnetic metals.

\acknowledgments

We thank the staff of the TRIUMF Center for Molecular and Materials Science
and our TWIST collaborators for their encouragement and support.
In particular, the assistance of B. Hitti, R. Abasalti, and D. Vyas is
gratefully acknowledged.
This work was supported in part by the Natural Sciences and Engineering 
Research Council and the National Research Council of Canada, the Russian 
Ministry of Science, and the U.S.\@ Department of Energy.

%


\begin{thebibliography}{49}
\expandafter\ifx\csname natexlab\endcsname\relax\def\natexlab#1{#1}\fi
\expandafter\ifx\csname bibnamefont\endcsname\relax
  \def\bibnamefont#1{#1}\fi
\expandafter\ifx\csname bibfnamefont\endcsname\relax
 \def\bibfnamefont#1{#1}\fi
 \expandafter\ifx\csname citenamefont\endcsname\relax
  \def\citenamefont#1{#1}\fi
\expandafter\ifx\csname url\endcsname\relax
  \def\url#1{\texttt{#1}}\fi
\expandafter\ifx\csname urlprefix\endcsname\relax\def\urlprefix{URL }\fi
\providecommand{\bibinfo}[2]{#2}
\providecommand{\eprint}[2][]{\url{#2}}

\bibitem{Brewer:1994}
J. H. Brewer, Muon spin rotation/relaxation/resonance, in {\it Encyclopedia of Applied Physics} (VCH, New York, 1994), Vol. 11, p. 23.
\bibitem{TWIST}
R. Bayes {\it et al.}, Phys. Rev. Lett. \textbf{106}, 041804 (2011).
\bibitem{Cox:1987}
S.F.J. Cox, J. Phys. C: Solid State Phys. \textbf{20}, 3187 (1987).
\bibitem{Schilling:1978}
W. Schilling, Hyperfine Interactions \textbf{4}, 636 (1978).
\bibitem{Brice:1978}
D.K. Brice, Phys. Lett. \textbf{66A}, 53 (1978).
\bibitem{Dalmas:1997} 
P. Dalmas de R\'{e}otier and A. Yaouanc, J.Phys.: Condens. Matter \textbf{9}, 9113 (1997).  
\bibitem{Abragam:1986}
A. Abragam, Principles of Nuclear Magnetism, in {\it International series of monographs on physics\/} (Oxford University Press, 1961).
\bibitem{Hayano:1979}
R.S.~Hayano, Y.J.~Uemura, J.~Imazato, N.~Nishida, T.~Yamazaki, and R~Kubo, Phys. Rev. B \textbf{20}, 850 (1979).
\bibitem{Dalmas:1992}
P. Dalmas de R\'{e}otier and A. Yaouanc, J.Phys: Condens. Matter \textbf{4}, 4533 (1992).
\bibitem{Slichter:1978}
C.P.~Slichter, {\it Principles of Magnetic Resonance} (Springer-Verlag, Berlin, 1978).
\bibitem{Richter:1986}
D.~Richter, Hyperfine Interactions \textbf{31}, 169 (1986).
\bibitem{Korringa:1950}
J.~Korringa, Physica \textbf{16}, 601 (1950).
\bibitem{Blundell:2001}
S.J. Blundell and S.F.J. Cox, J. Phys.: Condens. Matter \textbf{13}, 2163 (2001). 
\bibitem{Brewer:1975}
J.H. Brewer {\it et al.}, Positive Muons and Muonium in Matter, in {\it Muon Physics}, eds. Vernon W. Hughes and C.S. Wu (Academic Press, New York, 1975).
\bibitem{Cox:2000}
S.F.J.~Cox, S.P.~Cottrell, M.~Charlton,P.A.~Donnelly, S.J.~Blundell, J.L.~Smith, J.C.~Cooley, and W.L.~Hults, Physica B \textbf{289-290}, 594 (2000).  
\bibitem{Brown:1981}
J.A.~Brown {\it et al.}, Phys. Rev. Lett \textbf{47}, 261 (1981).
\bibitem{Blairpaper}
B.~Jamieson {\it et al.}, Phys. Rev. D \textbf{74}, 072007 (2006).
\bibitem{Heffner:1981}
R.H. Heffner, Hyperfine Interactions \textbf{8}, 655 (1981).
\bibitem{Kossler:1978} 
W.J. Kossler {\it et al.}, Phys. Rev. Lett. \textbf{41}, 1558 (1978). 
\bibitem{Hartmann:1978}
O. Hartmann, E~Karlsson, L.O.~Norlin, D.~Richter, and T.O.~Niinikoski, Phys. Rev. Lett. \textbf{41}, 1055 (1978). 
\bibitem{Kehr:1982}
K.W. Kehr {\it et al.}, Phys. Rev. B \textbf{26}, 567 (1982). 
\bibitem{Kehr:1984}
K.W. Kehr, Hyperfine Interactions \textbf{17-19}, 63 (1984). 
\bibitem{Hartmann:1988}
O. Hartmann {\it et al.}, Phys. Rev. B \textbf{37}, 4425 (1988). 
\bibitem{Hartmann:1990}
O. Hartmann, Hyperfine Interactions \textbf{64}, 641 (1990). 
\bibitem{Hartmann:1980}
O. Hartmann {\it et al.}, Phys. Rev. Lett. \textbf{44}, 337 (1980). 
\bibitem{Stoker:1985_2}
D.P. Stoker {\it et al.}, Phys. Rev. Lett. \textbf{54}, 1887 (1985). 
\bibitem{private_stoker}
D.P. Stoker, University of California, Irvine, USA. Private communication, 2009.  
\bibitem{Jodidio}
A.~Jodidio {\it et al.}, Phys. Rev. D \textbf{34}, 1967 (1986); \textbf{37}, 237(E) (1988).
\bibitem{Beveridge:1985}
J.~L.~Beveridge {\it et al.}, Nucl. Instrum. Methods A \textbf{240}, 316 (1985).\bibitem{Goodfellow}
Goodfellow Corporation, 305 High Tech Drive, Oakdale, PA 15071-3911, USA.
\bibitem{Gauster:1977}
W.B. Gauster {\it et al.}, Solid State Communications \textbf{24}, 619 (1977).
\bibitem{ESPI}
ESPI Metals, 1050 Benson Way, Ashland, Oregon 97520, USA.
\bibitem{Dunsiger}
S.R.~Dunsiger {\it et al.}, Phys. Rev. B \textbf{73}, 172418 (2006).
\bibitem{SRIM}
J.~F.~Ziegler, Nucl. Instrum. Methods B \textbf{219}, 1027 (2004).
\bibitem{Bueno}
J.F. Bueno {\it et al.}, Phys. Rev. D (to be submitted); J.F.~Bueno, PhD. Thesis, University of British Columbia, 2010 (http://hdl.handle.net/2429/23724).

\end{thebibliography}
\end{document}